\documentclass[universe,article,accept,moreauthors,pdftex,10pt,a4paper]{mdpi}

\firstpage{1} 
\pubvolume{7}
\articlenumber{461}
\doinum{10.3390/universe7120461}
\pubyear{2021}
\copyrightyear{2021}
\externaleditor{Academic Editor: Riccardo Brugnera}
\history{Received: 31 August 2021 /  Accepted: 15 November 2021 / Published: 25 November 2021}

\pdfoutput=1
\usepackage[T1]{fontenc} 
\usepackage[utf8]{inputenc}

\usepackage{graphicx}
\usepackage{graphics}
\usepackage[mathscr]{eucal}
\usepackage{amssymb} 
\usepackage{amsmath}
\usepackage{xspace}
\usepackage{listings}
\usepackage{ulem}
\usepackage{xcolor}
\usepackage{bm}         
\usepackage{array}
\usepackage{multirow}   
\usepackage{booktabs}   
\usepackage{mleftright}

\usepackage{bbold}
\usepackage{amsmath,amscd}
\usepackage{slashed}
\usepackage{placeins}
\usepackage{soul}
\usepackage{braket}
\usepackage{tikz-cd}
\usepackage{empheq}
\usepackage[most]{tcolorbox}
\usetikzlibrary{arrows, shapes,shadows}
\usepackage{mathtools,slashed}
\usepackage{dsfont}
\usepackage[makeroom]{cancel}
\usepackage{multirow}

\usepackage{amssymb}
\usepackage[capitalise]{cleveref}
 \theoremstyle{mdpi}
 \newcounter{thm}
 \newcounter{ex}
 \newcounter{re}
\newcommand{\vev}[0]{VEV\xspace}
\newcommand{\vevs}[0]{VEVs\xspace}

\newcommand*{\Scale}[2][4]{\scalebox{#1}{$#2$}}%

\newcommand{\U}[1]{\mathrm{U}(1)_{\mathrm{#1}}}			
\newcommand{\SU}[2]{\mathrm{SU}(#1)_{\mathrm{#2}}}		
\newcommand{\SO}[2]{\mathrm{SO}(#1)_{\mathrm{#2}}}		
\newcommand{\E}[1]{\mathrm{E}_{#1}}		

\newcommand{\LLR}[3]{\left(\bm{L}^{ #1} \right)^{ #2 }{}_{ #3 }}
\newcommand{\QL}[3]{\left(\bm{Q}_{\mathrm{L}}^{ #1} \right)^{ #2 }{}_{ #3 }}
\newcommand{\QR}[3]{\left(\bm{Q}_{\mathrm{R}}^{ #1} \right)^{ #2 }{}_{ #3 }}

\definecolor{bostonuniversityred}{rgb}{0.8, 0.0, 0.0}

\newcommand{\mean}[1]{\left \langle #1 \right \rangle }

\newcommand{\ro}{\textrm}

\Title{Grand Unified origin of gauge interactions and families replication
in the Standard Model}

\Author{Ant{\'o}nio~P.~Morais$^{1}$, Roman~Pasechnik$^{2}$ and Werner~Porod$^{3}$}
\address{$^{1}$ \quad 
Departamento de F\'isica, Universidade de Aveiro and CIDMA, Campus de Santiago,
3810-183 Aveiro, Portugal; aapmorais@ua.pt \\
$^{2}$ \quad
Department of Astronomy and Theoretical Physics, Lund
University, SE-223 62 Lund, Sweden; Roman.Pasechnik@thep.lu.se \\
$^{3}$ \quad
Institut für Theoretische Physik und Astrophysik, Uni Würzburg, Germany; porod@physik.uni-wuerzburg.de}

\corres{Correspondence: Roman.Pasechnik@thep.lu.se}

\abstract{The tremendous phenomenological success of the Standard Model (SM) suggests that its 
flavor structure and gauge interactions may not be arbitrary but should have a fundamental 
first-principle explanation. In this work, we explore how the basic distinctive properties 
of the SM dynamically emerge from a unified New Physics framework tying together both 
flavour physics and Grand Unified Theory (GUT) concepts. 
This framework is suggested by a novel anomaly-free supersymmetric chiral 
$\E{6} \times \SU{2}{F} \times \U{F}$ GUT containing the SM.
Among the most appealing emergent properties 
of this theory is the Higgs-matter 
unification with a highly-constrained massless chiral sector featuring two universal 
Yukawa couplings close to the GUT scale. At the electroweak scale, the minimal SM-like 
effective field theory limit of this GUT represents a specific flavored 
three-Higgs doublet model consistent with the observed large hierarchies in the quark 
mass spectra and mixing already at tree level. }


\begin{document}

With a handful of physical parameters such as fermion masses and gauge 
couplings the Standard Model (SM) explains a huge variety of collider 
and low energy data spanning over several orders of magnitude
for the corresponding energy scales. Its success builds strongly on 
the gauge principle. However, it is fundamentally incomplete as it leaves 
the cosmological Dark Matter and baryon asymmetry 
of the universe unexplained. Moreover, it neither contains mechanisms 
for generating the tiny neutrino masses 
nor explains the structure of the SM fermion families. This
suggests that the SM is not the ultimate theory but an excellent 
effective field theory (EFT) of the subatomic world.

Since the birth of the SM in mid-1970, there have been numerous attempts 
to come up with a consistent first-principle explanation of the well-measured 
but yet totally arbitrary and rather odd properties of the SM. Among these are 
the remarkable proton stability, the specific structure of gauge and Yukawa 
interactions and the properties of the Higgs and Yukawa sectors which are
intimately connected to the rather peculiar observed patterns in the neutrino 
and charged fermion mass spectra and generation mixings (the so-called flavor problem).

It is fairly easy to achieve unification of
the gauge couplings at higher energy scales by postulating
the existence of additional scalars and/or fermions 
belonging to incomplete representations of $\SU{5}{}$
\cite{Calibbi:2009cp}. This is for example realized
in supersymmetric (SUSY) extensions of the SM \cite{Martin:1997ns}.
This unification is a necessary requirement to embed the SM
into a larger gauge group such as as~$\SU{5}{}$, $\SO{10}{}$, or $\mathrm{E}_6$ \cite{Georgi:1974sy,Fritzsch:1974nn,Chanowitz:1977ye,Georgi:1978fu,
Georgi:1979dq,Georgi:1979ga,Georgi:1982jb,Gursey:1975ki,Gursey:1981kf,Achiman:1978vg,Pati:1974yy}, so-called grand unified theories (GUTs). For a recent thorough discussion of theoretical features and most important phenomenological implications 
of the $\mathrm{E}_6$ GUTs, see e.g.~Refs.~\citep{Athron:2008np,Hall:2010ix,Rizzo:2012rf,Nevzorov:2012hs,Nevzorov:2013tta,Nevzorov:2015sha,Athron:2015vxg,Ishihara:2015uua}. 
Remarkably, in Ref.~\cite{Berezhiani:1989bd} it has been demonstrated that the well-known hierarchy and doublet-triplet splitting problems appear to be naturally resolved in the framework of SUSY $\SU{6}{}$ GUT (see also Refs.~\cite{Berezhiani:1995sb,Dvali:1996sr}). This model also provides means for explanation of the origin of the fermion mass hierarchy, i.e. why the 3rd fermion family in the SM is heavier than first two \cite{Barbieri:1994kw}.
Other promising scenarios designed to address the flavor problem invoke new ``horizontal'' symmetries at high energies; for recent studies, see 
e.g.~Refs.~\cite{Ordell:2019zws,CarcamoHernandez:2019cbd,Vien:2019zhs,CarcamoHernandez:2019vih,
CarcamoHernandez:2018iel,Bjorkeroth:2019csz,Gui-JunDing:2019wap}. For original works, where it was suggested, in particular in the context of $\SU{3}{F}$, that the observed fermion spectrum with its hierarchies of masses and mixing angles are due to horizontal symmetry breaking hierarchy, i.e. by the flavon VEV hierarchy breaking it, see Refs.~\cite{Berezhiani:1983rk,Berezhiani:1985in,Berezhiani:1990wn,Berezhiani:1996ii}.

In general, it is rather difficult to combine both gauge symmetries and horizontal symmetries without a proliferation of unknown parameters in the Yukawa sector. Our main goal here is to find a consistent GUT framework in four spacetime dimensions, with both 
types of unification realised dynamically in the gauge and Yukawa sectors. For this purpose, 
we would like to study high-scale SUSY-based framework binding together both observed 
fermion families' replication in the SM and grand unification. For this purpose, let us consider
the ${\cal N}=1$ SUSY $\E{6} \times \SU{2}{F} \times \U{F}$ GUT in four dimensions where the gauge 
symmetries of the SM originate from $\E{6}$ whereas additional group factors $\SU{2}{F} \times \U{F}$ 
conveniently represent a ``horizontal'' gauge symmetry distinguishing the fermion families, i.e.~the family 
symmetry\footnote{For a previous discussion of implications of non-abelian family symmetries in supersymmetric 
GUT model building, see e.g. Refs.~\citep{Ross:2002mr,Ross:2004qn}}. 

Note that a further constrained scenario based on $\SU{3}{F}$ has been previously studied by some of the authors in Refs.~\cite{Camargo-Molina:2016yqm,Camargo-Molina:2017kxd}. Indeed, promoting $\SU{2}{F} \times \U{F}$ to $\SU{3}{F}$, the model seems to become more compact and natural. However, it was found that the top and charm quark tree-level masses are degenerate and a strong fine-tuning in the soft SUSY breaking sector is necessary in order to induce a realistic mass splitting at one-loop level. Such a fine-tuning implies a rather strong hierarchy between different soft SUSY breaking parameters that has prompted the search for a less constrained fully-gauged family symmetry such as the one considered in this work.

The subsequent symmetry breaking steps can be realised by means of the Higgs mechanism as follows:
\begin{eqnarray}
     \mkern-18mu\E{6} \times \SU{2}{F} \times \U{F} &\overset{M_6}{\longrightarrow}&~[\SU{3}{}]^3 \times \SU{2}{F} \times \U{F} \label{eq:E6SSB}\\
     &\overset{M_3}{\longrightarrow}&~\SU{3}{C} \times [\SU{2}{} \times \U{}]^2  \label{eq:Tri-SSB} \\ 
     &\qquad& \times \SU{2}{F}\times \U{F}
     \overset{M_\ro{S}}{\longrightarrow} \dots  \label{eq:Soft-SSB} \,.
\end{eqnarray}
where $[\SU{3}{}]^3\equiv \SU{3}{C} \times \SU{3}{L} \times \SU{3}{R}$ is the trinification group 
and $[\SU{2}{} \times \U{}]^2\equiv \SU{2}{L} \times \SU{2}{R}\times \U{L} \times \U{R}$.
We adopt at this stage that $\E{6} \times \SU{2}{F} \times \U{F}$ theory originates from a certain large gauge group ${\cal G}$ 
at the upper-most GUT-scale $M_{\rm GUT}$ (with a single universal gauge coupling) by means of some unknown 
dynamics and formulate the basic criteria for phenomenological consistency of such a scenario.

The mass scales of the rank-preserving symmetry breaking steps in Eqs.~(\ref{eq:E6SSB}) and (\ref{eq:Tri-SSB}) are given
by the sizes of the superpotential quadratic terms implying a nearly-compressed scale hierarchy
$M_{\rm GUT} \gtrsim M_6 \gtrsim M_3$. The $\dots$ in Eq.~(\ref{eq:Soft-SSB}) represent the subsequent low-scale 
breaking steps down to the SM gauge group triggered by soft-SUSY breaking interactions
at the soft scale $M_\ro{S}$. The latter can be decoupled from the trinification breaking scale, i.e.\ $M_\ro{S}\ll M_3$, 
in consistency with the low-scale electroweak spontaneous symmetry breaking (EW-SSB) in the SM.
\begin{table}[htb!]
	\begin{center}
		\begin{tabular}{c|c}
			\toprule                     
		         $\mathbb{Z}_2$-even   & $\mathbb{Z}_2$-odd  \\    
			\midrule
		       $\bm{\psi}^{\mu\,i} = \left(\bm{27},\bm{2} \right)_{(1)}\,, \quad \bm{\psi}^{\mu\,3} = \left(\bm{27},\bm{1} \right)_{(-2)}$  & \\
		       $\bm{{\cal H}}_{\cal U} = \left(\bm{1},\bm{2} \right)_{(-1)}\,, \quad \bm{{\cal H}}_{\cal D} = \left(\bm{1},\bm{2} \right)_{(+1)}$  
		       & $\bm{{\cal L}}_k = \left(\bm{1},\bm{2} \right)_{(-1)}$   \\ 
		       $\bm{{\cal A}} = \left( \bm{78},\bm{1} \right)_{(0)}$  & $\bm{{\cal E}}_k = \left(\bm{1},\bm{1} \right)_{(+2)}$  \\
		       $\bm{\Sigma}, \bm{\Sigma}^\prime = \left(\bm{650},\bm{1} \right)_{(0)} $ 
		       & $\bm{{\cal N}}_k = \left(\bm{1},\bm{1} \right)_{(0)} $  \\
		       $\bm{\Psi} = \left(\bm{2430},\bm{1} \right)_{(0)} $ &   \\
			\bottomrule
		\end{tabular} 
		\caption{Fundamental superfield content of the anomaly-free 4-dimensional SUSY $\E{6} \times \SU{2}{F} \times \U{F}$ GUT. 
		Here, $i=1,2$ and $k=1,2,3$.
		}
		\label{tab:Z4}  
	\end{center}
\end{table}

In this work, our main goal is to briefly discuss the main features of the SUSY $\E{6} \times \SU{2}{F} \times \U{F}$ GUT with the symmetry 
breaking pattern given in Eqs.~(\ref{eq:E6SSB}) - (\ref{eq:Soft-SSB}) and with a particular anomaly-free superfield content summarised 
in Table~\ref{tab:Z4}. In order to build a minimal working GUT scenario in this framework, one simple way to proceed is to adopt a scheme enabling 
to give large masses to the majority of supermultiplets listed in Table~\ref{tab:Z4} apart from those that we need for phenomenological 
explorations. For simplicity, we introduce an extra $\mathbb{Z}_2$ parity that plays a similar role to $R$-parity of the MSSM 
yielding the superpotential of the $\mathbb{Z}_2$-odd superfields analogical to that of the MSSM leptonic sector with right-handed neutrinos,
\begin{eqnarray}
W_{\bm{{\cal H}{\cal N}}} = \mu_{\cal H} \bm{{\cal H}}_{\cal U} \bm{{\cal H}}_{\cal D} + y_{\cal L} \bm{{\cal H}}_{\cal U} \bm{{\cal L}} \bm{{\cal E}}
+ y_{\cal N} \bm{{\cal H}}_{\cal D} \bm{{\cal L}} \bm{{\cal N}} + \mu_{\cal N}\bm{{\cal N}}\bm{{\cal N}} \,.
\end{eqnarray}
Hence, an analogue of right-handed neutrino, $\bm{{\cal N}}$, receives Majorana mass at some large scale $\mu_{\cal N}$, while 
other additional superfields $\bm{{\cal H}}_{{\cal U},{\cal D}}$, $\bm{{\cal L}}$ and $\bm{{\cal E}}$ acquire their 
masses upon breaking of $\SU{2}{F} \times \U{F}$ symmetry through VEVs in scalar components of $\SU{2}{F}$ $\bm{{\cal H}}_{{\cal U},{\cal D}}$ doublets.
This resembles the Higgsino, (s)neutrinos' and (s)leptons' mass generation via the EW symmetry breaking mechanism in the conventional MSSM framework, but 
for the family $\SU{2}{F} \times \U{F}$ SSB and at a larger scale, $\mu_{\cal H} \gtrsim M_\ro{S} \gg M_{\rm EW}$. 

Provided that $\bm{{\cal H}}_{{\cal U},{\cal D}}$ are $\mathbb{Z}_2$-even, their \vevs are not affecting $\mathbb{Z}_2$ symmetry which is therefore preserved 
in this model and survives down to low scales. Similarly to $R$-parity, this symmetry provides a possible way to stabilise the lightest state among the $\mathbb{Z}_2$-odd 
components of $\bm{{\cal L}}$, $\bm{{\cal E}}$ and $\bm{{\cal N}}$ superfields. Whether or not such a state can play a role of a Dark Matter candidate remains 
one of the interesting topics for further studies in this model.
Besides, neutrino-like states may receive a relatively small mass scale due to a seesaw-type mixing with the Majorana fermion from $\bm{{\cal N}}$.
Communication of such $\mathbb{Z}_2$-odd sector with the SM sectors would be suppressed due to a large mass scale of family 
$\SU{2}{F} \times \U{F}$ gauge bosons at tree level and due to a small loop-generated coupling to the SM Higgs boson. Apart from potentially light and 
decoupled $\mathbb{Z}_2$-odd states, all the other fields in $W_{\bm{{\cal H}{\cal N}}}$ are well above the EW scale and their impact on the low-scale 
phenomenology is expected to be strongly suppressed. Thus, they can be safely integrated out below $M_\ro{S}$ scale.

It is worth noticing here that the particle content of the considered $\E{6} \times \SU{2}{F} \times \U{F}\times \mathbb{Z}_2$ GUT 
and charge assignments in Table~\ref{tab:Z4} enable the superpotential mass terms to all the fundamental $\mathbb{Z}_2$-even superfields except for $\bm{\psi}^{\mu\,i}$ and 
$\bm{\psi}^{\mu\,3}$, providing a novel GUT framework manifestly free of the gauge and Witten anomalies.

The large $\E{6}$ representations $\bm{\Sigma}, \bm{\Sigma}^\prime$ and $\bm{\Psi}$ trigger (through their scalar \vevs)  the spontaneous 
rank- and SUSY-preserving breaking of $\E{6}$ symmetry at $M_6$ scale in Eq.~(\ref{eq:E6SSB}), while the components of the $\mathbb{Z}_2$-even $\E{6}$-adjoint 
representation $\bm{{\cal A}}$ play a critical role in triggering the subsequent trinification SSB in Eq.~(\ref{eq:Tri-SSB}). All these fields conveniently
received large masses and can be integrated out below either $M_6$ (large $\E{6}$ reps) or $M_3$ (adjoint $\E{6}$ rep) scale in the considered SUSY GUT. 
Taking the trinification breaking at $M_3$ scale as an example, the $\SU{3}{L,R}$-adjoint superfields $\bm{\Delta}^a_{\rm L,R}\subset \left(\bm{78},\bm{1} \right)$ 
(see Table~\ref{tab:Superfields}) originating from $\bm{{\cal A}}$ superfield have a universal mass term $\mu_{\bm{78}} \sim M_{\rm GUT}$ in the superpotential, which 
together with a cubic term triggers a rank- and SUSY-preserving \vev in one of its scalar components $\langle\tilde\Delta^{a=8}_{\rm L,R}\rangle\equiv M_{\rm 3}$ 
\cite{Camargo-Molina:2017kxd}. The SUSY-preserving breaking of a gauge symmetry implies that the $D$- and $F$-terms have to vanish separately. 
This means that the scalar potential has zero value in both the $\SU{3}{L,R}$-symmetric and $\SU{3}{L,R}$-broken vacua in the exact SUSY case. Thus, 
the presence of even a tiny soft-SUSY breaking effect already at the $M_{\rm GUT}$ scale is needed to make these vacua non-equivalent \cite{Witten:1981nf}, 
hence, enabling the trinification SSB in Eq.~(\ref{eq:Tri-SSB}). As a result, all the components of $\bm{\Delta}^a_{\rm L,R}$ acquire a universal 
mass, $M_{\Delta_{\ro{L,R}}}\sim M_{3}$ due to $D$-terms and thus are integrated out below the $M_{\Delta_{\ro{L,R}}}\sim M_{\rm 3}$ scale.

The remaining massless superfields $\bm{\psi}^{\mu\,i}$ and $\bm{\psi}^{\mu\,3}$ neatly unite 
the SM Higgs and matter (neutrino, charged lepton and quark) sectors hence imposing very specific constraints on the structure of the resulting low-energy EFT 
below $M_{\rm S}$. How does such Higgs-matter unification comply with observations? Notably, the considered $\E{6} \times \SU{2}{F} \times \U{F}$ SUSY GUT 
enables for a phenomenologically consistent splitting between the second- and third-generation quark masses already at tree level with only two distinct quark 
Yukawa couplings below $M_6$. Let us explore this interesting phenomenon in more detail.
\begin{table}[htb!]
	\begin{center}
		\begin{tabular}{cccccc|cc}
			\toprule                     
		            	            & $\SU{3}{L}$ & $\SU{3}{R}$    & $\SU{3}{C}$  & $\SU{2}{F}$ & $\U{F}$ & $\U{W}$ & $\U{B}$ \\    
			\midrule
			$\bm{L}^{i}$            & $\bm{3}$ & $\bm{\overline{3}}$ & $\bm{1}$  & $\bm{2}$	& $1$	& $1$	& $0$	\\
			$\bm{L}^{3}$            & $\bm{3}$ & $\bm{\overline{3}}$ & $\bm{1}$  & $\bm{1}$	& $-2$	& $1$	& $0$	\\
			$\bm{Q}^{i}_\mathrm{L}$ & $\bm{\overline{3}}$ & $\bm{1}$ & $\bm{3}$  & $\bm{2}$	& $1$	& $-1/2$	& $1/3$   \\
			$\bm{Q}^{3}_\mathrm{L}$ & $\bm{\overline{3}}$ & $\bm{1}$ & $\bm{3}$  & $\bm{1}$	& $-2$	& $-1/2$	& $1/3$   \\
			$\bm{Q}^{i}_\mathrm{R}$ & $\bm{1}$ & $\bm{3}$ & $\bm{\overline{3}}$  & $\bm{2}$	& $1$	& $-1/2$	& $-1/3$   \\
			$\bm{Q}^{3}_\mathrm{R}$ & $\bm{1}$ & $\bm{3}$ & $\bm{\overline{3}}$  & $\bm{1}$	& $-2$	& $-1/2$	& $-1/3$   \\
			\midrule
			$\bm{\Delta}_\mathrm{L}$ & $\bm{8}$ & $\bm{1}$ & $\bm{1}$            & $\bm{1}$	& $0$	& $0$	& $0$	\\
			$\bm{\Delta}_\mathrm{R}$ & $\bm{1}$ & $\bm{8}$ & $\bm{1}$            & $\bm{1}$	& $0$	& $0$	& $0$   \\
			$\bm{\Delta}_\mathrm{C}$ & $\bm{1}$ & $\bm{1}$ & $\bm{8}$            & $\bm{1}$	& $0$	& $0$	& $0$   \\
			$\bm{\Xi}$               & $\bm{3}$ & $\bm{\overline{3}}$ & $\bm{3}$ & $\bm{1}$	& $0$	& $0$	& $0$   \\
			$\bm{\Xi^\prime}$  & $\bm{\overline{3}}$ & $\bm{3}$ & $\bm{\overline{3}}$ & $\bm{1}$ & $0$ & $0$ & $0$ \\
			\bottomrule
		\end{tabular} 
		\caption{Upper part: fundamental chiral superfields in the $[\SU{3}{}]^3 \times \SU{2}{F} \times \U{F}$ 
		theory -- components of the massless $\bm{\psi}^{\mu\,i}$ and $\bm{\psi}^{\mu\,3}$ superfields of $\E{6} \times \SU{2}{F} \times \U{F}$ \cite{Slansky:1981yr}. 
		Lower part: the corresponding components of the massive superfield $(\bm{78},\bm{1})_0$. 
		Accidental symmetries' charges are shown in last two columns.}
		\label{tab:Superfields}  
	\end{center}
\end{table}

First of all, the SUSY $\E{6} \times \SU{2}{F} \times \U{F}$ theory below $M_{\rm GUT}$ features a vanishing dim-3 superpotential 
$d_{\mu \nu \lambda} \varepsilon_{ij} \bm{\psi}^{\mu\,i} \bm{\psi}^{\nu\,j} \bm{\psi}^{\lambda\,3} = 0$,
caused by anti-symmetry of family index contractions, where $d_{\mu \nu \lambda}$ is a completely symmetric $\E{6}$ tensor 
\cite{Kephart:1981gf,Deppisch:2016xlp}, and $\varepsilon_{ij}$ is the totally anti-symmetric $\SU{2}{}$ Levi-Civita pseudotensor. 
Since the renormalisable $\E{6}$ interactions cannot generate a non-trivial Yukawa structure in this theory, the effects of 
high-dimensional operators become important. In particular, the relevant part of the superpotential below $M_{\rm GUT}\gtrsim M_6$ 
scale that contains the SM matter and Higgs sectors reads
\begin{eqnarray}    
\label{eq:super4D}
   && W_{\bm{\psi}} = \frac{\varepsilon_{ij} \bm{\psi}^{\mu\,i} \bm{\psi}^{\nu\,j} \bm{\psi}^{\lambda\,3}}{2M_{\rm GUT}} 
   \Big[ \tilde{\lambda}_1 \bm{\Sigma}^\alpha_\mu d_{\alpha \nu \lambda} + 
   \tilde{\lambda}_2 \bm{\Sigma}^\alpha_\nu d_{\alpha \mu \lambda} +
   \tilde{\lambda}_4 {\bm{\Sigma^\prime}}^\alpha_\mu d_{\alpha \nu \lambda} + 
   \tilde{\lambda}_5 {\bm{\Sigma^\prime}}^\alpha_\nu d_{\alpha \mu \lambda} 
   \Big] \,,  
\end{eqnarray}
where the minimal superfield content necessary to generate two distinct Yukawa couplings 
below $M_6$ requires the presence of two different bi-fundamental $\bm{650}$-superfields 
of $\E{6}$ \cite{Gursey:1980uj}, $\bm{\Sigma}^\mu_\nu$ and ${\bm{\Sigma^\prime}}^\mu_\nu$. Their \vevs, 
$\mean{\bm{\Sigma}} ~\propto~ k_\Sigma M_6$ and $\mean{\bm{\Sigma^\prime}} 
~\propto~ k_{\Sigma^\prime} M_6$, trigger subsequent breaking of $\E{6}$ 
\cite{Chakrabortty:2008zk} down to the trinification symmetry (\ref{eq:E6SSB}). 
As a result, an EFT superpotential of massless fields below $M_6$ reads
\begin{eqnarray}
  	W_{\rm eff} =\Scale[0.96]{\varepsilon_{ij} (
  	\mathcal{Y}_{1} \bm{L}^i \cdot \bm{Q}^3_\mathrm{L} \cdot \bm{Q}^j_\mathrm{R}
  	- \mathcal{Y}_{2} \bm{L}^i \cdot \bm{Q}^j_\mathrm{L} \cdot \bm{Q}^3_\mathrm{R} 
  	+ \mathcal{Y}_{2}\bm{L}^3 \cdot \bm{Q}^i_\mathrm{L}\cdot \bm{Q}^j_\mathrm{R} )}
  	\label{super2}
\end{eqnarray}
in terms of the massless trinification leptonic $\bm{L}^{i,3}$ and quark $\bm{Q}^{i,3}_\mathrm{L,R}$ 
superfields -- components of the original massless $\bm{\psi}^{\mu\,i}$ and $\bm{\psi}^{\mu\,3}$ 
superfields of $\E{6} \times \SU{2}{F} \times \U{F}$ described in Table~\ref{tab:Superfields}. Next, it is convenient 
to perform the following decomposition
\begin{equation}
\Scale[0.95]{\LLR{i,3}{l}{r} =\begin{pmatrix}
	\bm{\chi}^{\bar{l}}{}_{\bar{r}} & \bm{\ell}^{\bar{l}}_{\mathrm{L}}\\
	\bm{\ell}_{\mathrm{R}\bar{r}} & \bm{\phi}
	\end{pmatrix}^{i,3}\,,} \quad
\begin{aligned}
&\Scale[0.95]{\QL{i,3}{x}{l}=\begin{pmatrix}\bm{q}_{\mathrm{L}\bar{l}}^x & \bm{D}_{\mathrm{L}}^x
	\end{pmatrix}^{i,3}},
\label{eq:tri-triplets}
\\
&\Scale[0.95]{\QR{i,3}{r}{x}=\begin{pmatrix}\bm{q}_{\mathrm{R}x}^{\bar{r}} & \bm{D}_{\mathrm{R}x}
	\end{pmatrix}^{\top\;\;i,3}},
\end{aligned}
\end{equation}
such that, upon further splitting into $\SU{2}{L,R}$ representations, one has
\begin{equation}
    \begin{aligned}
    \bm{\chi}^{\bar{l}}{}_{\bar{r}} = \begin{pmatrix}\bm{H}_{\mathrm{u}}^{\bar{l}} & \bm{H}^{\bar{l}}_{\mathrm{d}}
	\end{pmatrix}^{i,3}
	\quad
	\bm{\ell}_{\mathrm{R}\bar{r}}  =
	\begin{pmatrix} \bm{e}_{\mathrm{R}}  & \bm{\nu}_{\mathrm{R}}
	\end{pmatrix}^{i,3}
	\quad
	\bm{q}_{\mathrm{R}x}^{\bar{r}} =
	\begin{pmatrix} \bm{u}_{\mathrm{R}x}  & \bm{d}_{\mathrm{R}x}
	\end{pmatrix}^{\top \; \; i,3}\,.
    \end{aligned}
    \label{eq:SplitSU2}
\end{equation}
Above, $l$, $r$ and $x$ represent $\SU{3}{L}$, $\SU{3}{R}$ and $\SU{3}{C}$ triplet indices, $\bar{l}$ and $\bar{r}$ denote $\SU{2}{L}$ and $\SU{2}{R}$ doublet indices,
respectively, $i$ is the $\SU{2}{F}$ index, while the labels $\ro{L}$ ($\ro{R}$) should not be identified 
with left (right) chiralities at this stage (all fermionic components are L-handed Weyl spinors). In \cref{eq:SplitSU2} one can see that the model offers three up specific $\bm{H}^{i,3}_{\mathrm{u}}$ and three down specific $\bm{H}^{i,3}_{\mathrm{d}}$ Higgs doublet candidates. While by no means unique, and in fact not a preferred scenario according to the discussion in \cite{Morais:2020ypd}, a MSSM-like Higgs sector is a possible low-scale limit of the model. However, due to a non-trivial mixing structure, such doublets result from a linear combination of those in \cref{eq:SplitSU2} and cannot be promptly identified at the level of unbroken trinification. The Higgs-matter 
unification here implies that the Higgs doublet superfields of 
the EW theory are unified together with the lepton and quark $\SU{3}{L,R}$-superfields 
in the $\bm{\psi}^{\mu\,i}$ and $\bm{\psi}^{\mu\,3}$. Such a unification is thus enforced by the gauge
symmetry of the high-scale theory and that cannot be consistently realised in the MSSM.

The effective trinification superpotential (\ref{super2}) contains two universal Yukawa couplings
\begin{eqnarray}
    \mathcal{Y}_1 = \zeta \frac{k_{\Sigma^\prime}}{\sqrt{6}} \tilde{\lambda}_{45}\,, \quad
    \mathcal{Y}_2 = \zeta \frac{k_\Sigma}{2 \sqrt{2}} (\tilde{\lambda}_{21} - \tilde{\lambda}_{45})\,,
    \label{Y1Y2}
\end{eqnarray}
where $\tilde{\lambda}_{ij} \equiv \tilde{\lambda}_i - \tilde{\lambda}_j$ and 
$\zeta \simeq M_6/M_{\rm GUT}$. As we will demonstrate below, due to a very steep Renormalisation 
Group (RG) evolution of the gauge couplings in the $\E{6}\times \SU{2}{F} \times \U{F}$ theory 
at high scales and the required matching of the SM gauge couplings to their measured values 
at the electroweak (EW) scale, one has $\zeta \sim 1$ and $k_\Sigma \simeq - k_{\Sigma^\prime}$. 
On another hand, a possible common origin of the dim-4 operators from yet unknown $M_{\rm GUT}$-scale 
dynamics in the superpotential (\ref{eq:super4D}) and a compressed hierarchy $M_{\rm GUT} \gtrsim M_6$ 
imply that $\tilde{\lambda}_{21} \simeq \tilde{\lambda}_{45}$ suggesting the following hierarchy $\mathcal{Y}_2 \ll \mathcal{Y}_1 \sim 1$. 
It turns out that such an emergent hierarchy is consistent with the existence of an order-one top-quark 
Yukawa coupling given by $\mathcal{Y}_1$. Besides, it leads to the observed top-charm and bottom-strange quark mass hierarchies in the SM as well as to the down-type vector-like quark mass hierarchy already at tree level, namely,
\begin{equation}
\label{eq:ratio}
	\frac{\mathcal{Y}_1}{\mathcal{Y}_2} ~=~ \frac{m_\ro{t}}{m_\ro{c}} ~\approx~ \frac{m_\ro{b}}{m_\ro{s}}  ~\approx~ \frac{m_\ro{B}}{m_\ro{D,S}} ~\sim~ \mathcal{O}(100) \,,
\end{equation}
implying also a possibility for two light vector-like $D,S$-quark species potentially 
within the reach of the LHC or future collider measurements.

The superpotential \eqref{super2} possesses an accidental Abelian $\U{W} \times \U{B}$ 
symmetry whose charges, $W$ and $B$, are summarised in Table~\ref{tab:Superfields}. Furthermore, the theory has
an extra $\mathbb{Z}_2$ parity denoted as $\mathbb{P}_{\ro{B}}$-parity defined as
\begin{eqnarray}
    \mathbb{P}_{\ro{B}} &=& (-1)^{2 W + 2 S} = (-1)^{3 B + 2 S} \,,    
\end{eqnarray}
where $S$ is the spin. In the considered GUT theory, the $\mathbb{P}_{\ro{B}}$-parity replaces the conventional R-parity and forbids triple-squark or quark-quark-squark trilinear interactions in the soft-SUSY breaking sector capable of destabilising the proton at the soft scale. Together with the baryon-number $\U{B}$-symmetric Yukawa sector, this ensures that only $\E{6}$ gauge interactions can trigger the proton decay, highly suppressed by a large $M_6$ close to $M_{\rm GUT}$.

The dim-3 superpotential of $\bm{\Xi}$, $\bm{\Xi^\prime}$ and $\bm{\Delta}_\mathrm{L,R,C}$ superfields -- components of the massive chiral $\bm{{\cal A}}$ superfield (see Table~\ref{tab:Superfields}, also Ref.~\cite{Slansky:1981yr}) -- reads
\begin{equation}
    \begin{aligned}
    W_{\bm{78}} &= \sum_{A = \mathrm{L,R,C}} \Big[ \frac{1}{2} \mu_{78} \mathrm{Tr} \bm{\Delta}^2_A + \frac{1}{3!} \mathcal{Y}_{78} \mathrm{Tr} \bm{\Delta}^3_A \Big] 
    +\mu_{78} \mathrm{Tr} (\bm{\Xi} \bm{\Xi}^\prime) + \sum_{A = \mathrm{L,R,C}} \mathcal{Y}_{78} \mathrm{Tr}(\bm{\Xi} \bm{\Xi}^\prime \bm{\Delta}_A)\,,
    \end{aligned}
\end{equation}
with the universal $\mu_{78}\simeq M_{\rm GUT}$. As was mentioned above, the last rank/SUSY-preserving breaking step in Eq.~(\ref{eq:Tri-SSB}) represents the trinification breaking by means of degenerate \vevs at $M_3\lesssim M_6$ in the $\SU{3}{L}$, $\SU{3}{R}$ octet superfields $\bm{\Delta}_\mathrm{L}$, $\bm{\Delta}_\mathrm{R}$, respectively \cite{Camargo-Molina:2016yqm,Camargo-Molina:2017kxd}. In this case, all the $\bm{\Delta}_\mathrm{L,R,C}$ components acquire large masses $M_{\Delta_{\ro{L,R,C}}}\sim M_3$  and hence are integrated out
leaving no heavy fields in the resulting left-right (LR) symmetric SUSY EFT \cite{Morais:2020ypd}
\begin{equation}
\begin{aligned}
W_{\rm LR} & = \mathcal{Y}_{1} \varepsilon_{ij}\big[ \bm{\chi}^i \cdot \bm{q}^3_\mathrm{L}\cdot \bm{q}^{j}_\mathrm{R} 
+ \bm{\ell}^{i}_\mathrm{R} \cdot \bm{D}^3_\mathrm{L} \cdot \bm{q}^{j}_\mathrm{R} 
+ \bm{\ell}^i_\mathrm{L} \cdot \bm{q}^3_\mathrm{L} \cdot \bm{D}^{j}_\mathrm{R} 
+ \bm{\phi}^i \cdot \bm{D}^3_\mathrm{L} \cdot \bm{D}^{j}_\mathrm{R} \big]
\\ 
& - \mathcal{Y}_{2} \varepsilon_{ij}\big[j\leftrightarrow 3\big] 
+ \mathcal{Y}_{2} \varepsilon_{ij}\big[i\leftrightarrow 3\big]
\end{aligned}
\label{W_LR-SUSY}
\end{equation}
written in terms of the massless components of trinification bi-triplets introduced 
in Eq.~(\ref{eq:tri-triplets}). The further symmetry breaking steps down to the SM and 
hence the masses/mixings of the $\bm{L}^{i,3}$ and $\bm{Q}^{i,3}_\mathrm{L,R}$ components
are controlled by the structure of the soft-SUSY breaking mass terms 
and tri-linear interactions as well as by the tree-level Yukawa hierarchy (\ref{eq:ratio}). 

In the LR symmetric SUSY theory the largest amount of free parameters comes from 
the soft-SUSY breaking sector, namely, 17 trilinear couplings (5 invloving sleptons and 12 -- squarks),
16 soft $\tilde{L}\tilde{L}$- 
and $\tilde{Q}\tilde{Q}$-type mass terms, 2 high-scale gaugino mass parameters 
(in $E_6$ and gauge-family sectors). In addition, there are 4 gauge 
couplings in the gauge sector whereas all the low-scale Yukawa couplings are matched 
to two universal high-scale ones defining the strongest hierarchies (\ref{eq:ratio}) 
already at tree level. The loop corrections to the Yukawa sector are controlled 
by the soft-SUSY breaking parameters and gauge couplings,
whose number is sufficient to accommodate the measured
values of the SM fermion masses and mixing angles.

Let us now investigate how strong the hierarchy between the soft and trinification 
breaking scales, $M_{\rm S}\ll M_3$, can be -- the question of primary importance 
for a realistic low-energy theory. Provided the compressed hierarchy $M_{\rm GUT}\gtrsim M_6$, 
the unknown $M_{\rm GUT}$-scale effects may induce significant threshold corrections to the trinification 
gauge couplings at $M_6$ scale. Indeed, the relevant gauge-kinetic dim-5 operators 
\cite{Chakrabortty:2008zk} 
\begin{equation}
	\mathcal{L}_{\rm 5D} = -\frac{\xi}{M_{\rm GUT}} \Big[\frac{1}{4C}\, 
	{\rm Tr}(\bm{F}_{\mu \nu}\cdot \tilde{\Phi}_{\E{6}} \cdot \bm{F}^{\mu \nu})\Big] 
	\label{eq:5D}
\end{equation}
where $C$ is the charge normalization, $\bm{F}_{\mu \nu}$ is the $\E{6}$ field strength tensor, 
$\xi\sim 1$ is a non-renormalisable coupling constant, and $\tilde{\Phi}_{\E{6}}$ is a linear combination of 
the scalar fields originating from the symmetric product of two $\E{6}$ adjoint representations
	$\tilde{\Phi}_{\E{6}} \in ( \bm{78} \otimes \bm{78} )_\mathrm{sym} = 
	\bm{1} \oplus \bm{650} \oplus \bm{2430}$, with two $\bm{650}$-reps $\bm{\Sigma}^\mu_\nu$ and ${\bm{\Sigma^\prime}}^\mu_\nu$ 
already utilised above. The $\E{6}$-breaking \vevs in these 
fields modify the gauge coupling unification condition at $M_6$ scale via dim-5 
threshold corrections from Eq.~(\ref{eq:5D}) \cite{Chakrabortty:2008zk}
 \begin{equation}
    \begin{aligned}
    &\alpha_{3\mathrm{C}}^{-1} (1+ \zeta \delta_\mathrm{C})^{-1} = \alpha_{3\mathrm{L}}^{-1} 
 (1+ \zeta \delta_\mathrm{L})^{-1} = \alpha_{3\mathrm{R}}^{-1} (1+ \zeta \delta_\mathrm{R})^{-1} \,, \\
 &\alpha^{-1}_{3A} = \frac{4 \pi}{g_A^2} \,, \qquad \delta_\mathrm{C} 
 = - \frac{1}{\sqrt{2}}k_{\Sigma} - \frac{1}{\sqrt{26}} k_{\Psi} \,,
 \\
 &\delta_\mathrm{L,R} = \frac{1}{2\sqrt{2}}k_{\Sigma} \pm \frac{3}{2\sqrt{2}}
 k_{\Sigma^\prime} - \frac{1}{\sqrt{26}} k_{\Psi}\,, \quad 
 k_\Psi \propto \frac{\langle\bm{2430} \rangle}{M_6} \,,
 \end{aligned} 
 \label{eq:mod} 
 \end{equation}
where $\alpha^{-1}_{3A}$, $A=\mathrm{L},\mathrm{R},\mathrm{C}$, are the inverse trinification 
structure constants, $k_{\Sigma,\Sigma^\prime}$ and $\zeta\sim 1$ were defined above.
Provided that the family $\U{T}$ and the hypercharge $\U{Y}$ gauge groups remain unbroken above the EW scale, 
their the $T$- and $Y$-charges are related to the high-scale ones as $T_\ro{T} = 6 T_\ro{R}^3 - 4 T_\ro{F}^3 
+ \tfrac{2}{\sqrt{3}} ( T_\ro{L}^8 - T_\ro{R}^8 - 2 T_\ro{F}^8 )$ and $T_\ro{Y} = 2T_\ro{R}^3 + \tfrac{2}{\sqrt{3}} 
(T_\ro{L}^8 + T_\ro{R}^8)$, respectively. The corresponding inverse structure constants are matched (at tree level) 
to the high-scale ones below $M_3$-scale as follows: $\alpha_\ro{T}^{-1} = \tfrac{4}{9} \big(\alpha_\ro{2F}^{-1} +
\tfrac{1}{12}[\alpha_\ro{L}^{-1} + \alpha_\ro{R}^{-1} + 4 \alpha_\ro{F}^{-1}]\big) + \alpha_\ro{2R}^{-1}$ and 
$\alpha_\ro{Y}^{-1} = \tfrac{1}{3}(\alpha_\ro{L}^{-1} + \alpha_\ro{R}^{-1}) + \alpha_\ro{2R}^{-1}$, respectively. 
(Here, $\alpha_\ro{2A}$ and $\alpha_\ro{A}$ are the structure constants for $\SU{2}{A}$ and $\U{A}$, 
respectively.)

\begin{figure}[!htb]
		\begin{center}
		\includegraphics[width=.8\textwidth]{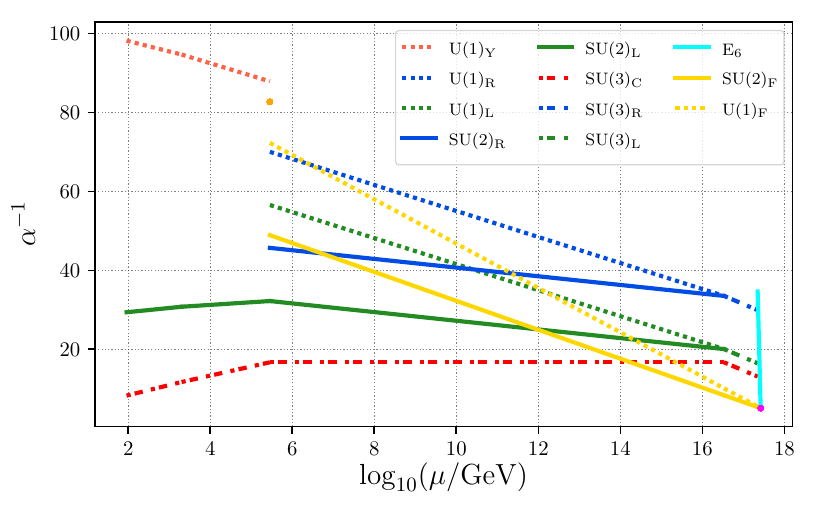}
		\end{center}
		\caption{An illustration of RG evolution of the gauge couplings in the considered model 
		for a parameter space point with a low-scale soft-SUSY breaking consistent 
		with the SM. Here, all the soft-induced symmetry breaking scales 
		(except $M_{\rm EW}\ll M_\ro{S}$) are assumed to be compressed for simplicity 
		and are fixed to a universal value $M_{\rm S}$ where the SM gauge symmetry 
		emerges. The purple dot represents the universal $\alpha^{-1}_{\rm U}(M_{\rm GUT})$, 
		while the orange dot shows the $\alpha^{-1}_\ro{T}(M_\ro{S})$ value. The purple dot represents an example of a unified gauge couplings where non-perturbative effects can start playing role.
}\label{fig:uni}
	\end{figure}

We have performed a sophisticated numerical analysis of the one-loop RG flow of gauge couplings 
between $M_{\rm GUT}$ and $M_{\rm EW}$ scales accounting for tree-level matching at intermediate scales 
as well as the matching to their measured counterparts at $M_{\rm EW}$. We have demonstrated that 
the presence of threshold corrections $\delta_A$ to the gauge couplings at $M_6$ 
enables a perturbative universal coupling with $M_{\rm GUT}=10^{16}-10^{18}$~GeV as well as 
low-scale soft-SUSY breaking down to as low as $M_{\ro{S}}\lesssim 10^3$ TeV,
in overall consistency with the SM phenomenology. For all valid points, we have found a compressed 
$M_6-M_{\rm GUT}$ hierarchy with $\zeta \simeq 1$ as well as $k_\Sigma \simeq - k_{\Sigma^\prime}$. 
One particular example for such RG flow for a valid parameter space point is shown in Fig.~\ref{fig:uni}.  The $\E{6}$ 
gauge coupling evolves very fast as indicated by a steep line stretched between $M_{\rm GUT}$ and $M_6$ scales.
The threshold corrections are quite sizable in this example. We give the corresponding parameters in \cref{tab:bench}.
Note, in an unrealistic case of a strong $M_6-M_{\rm GUT}$ hierarchy $\zeta\ll 1$, one could recover an approximate 
unification $\alpha_{3\mathrm{C}}^{-1} \simeq \alpha_{3\mathrm{L}}^{-1} \simeq \alpha_{3\mathrm{R}}^{-1}$ 
corresponding to $\mathbb{Z}_3$-permutation symmetry in the trinification gauge sector, 
originally imposed in Ref.~\cite{original}. However, a small $\zeta\ll 1$ implies unacceptably 
small Yukawa couplings (see Eq.~\eqref{Y1Y2}).
%
\begin{table}[t]
\begin{center}
\resizebox{\columnwidth}{!}{%
\begin{tabular}{ccccccccccccc}
    \toprule
                        $t_8$ & $t_3$ & $t_\ro{S}$ & $\zeta$ & $\alpha_8^{-1}\left(M_8\right)$ & $\alpha_\ro{T}^{-1}\left(M_\ro{S}\right)$ & $\delta_\ro{L}$ & $\delta_\ro{R}$ & $\delta_\ro{C}$ & $k_\Psi$ & $k_\Sigma$ & $k_\Sigma^\prime$ & $k_\sigma$ \\
                        \midrule
                        $17.42$ & $16.53$ & $5.455$ & $0.844$ & $5.067$ & $82.70$ & $-0.622$ & $-0.161$ & $-0.737$ & $0.862$ & $0.326$ & $-0.377$ & $0.0998$ \\
                        \bottomrule
                \end{tabular}}
                \caption{Benchmark points used for the
                running of the gauge couplings in \cref{fig:uni}. The top line corresponds to a parameter space point where $\delta_i$ differ considerably whereas in the bottom
                line their absolute values are of the same order. Here, $t_i = \log_{10}\tfrac{M_i}{\ro{GeV}}$.}
                \label{tab:bench}
        \end{center}
\end{table}

A simple configuration of soft-scale induced \vevs breaking the symmetry in Eq.~(\ref{eq:Soft-SSB})
down to $\SU{3}{C}\times \U{E.M.}$ reads
\begin{equation} 
\begin{aligned}
\mean{\tilde{L}^k} ~=~ \frac{1}{\sqrt{2}}\begin{pmatrix}
	{u_k} & 0 & 0\\
	0 & {d_k} & 0\\
	0 & \sim M_{\ro{S}} & \sim M_{\ro{S}}
	\end{pmatrix}\,, \qquad k=1,2,3 \,,
\end{aligned} 
\end{equation}
where $u_k$ and $d_k$ represent the EW-breaking Higgs up-type and down-type \vevs in the low-energy 
SM-like EFT, respectively, while all other \vevs are considered to be of order $M_{\ro{S}}$, 
for simplicity. With such a \vev setting, we can have up to six Higgs doublets in the low-energy
limit of the theory. However, for the sake of simplicity, here we consider that only five Higgs 
doublets acquire VEVs. In this case we found that there is only one particular configuration 
of these VEVs with $d_1=0$ that provides the Cabibbo–Kobayashi–Maskawa (CKM) matrix and the quark mass spectrum compatible with those in the SM, i.e.
\begin{eqnarray}
    V_{\rm CKM} =
\begin{pmatrix}
	 \frac{d_2 u_2 \mathcal{Y}_1^2 + d_3 u_3 \mathcal{Y}_2^2}{\sqrt{\mathcal{A}\mathcal{B}}} & -\frac{u_1 \mathcal{Y}_1 }{\sqrt{\mathcal{A}} } & 
	 \frac{(d_2 u_3 - d_3 u_2)\mathcal{Y}_1 \mathcal{Y}_2 }{\sqrt{\mathcal{A}\mathcal{B}}} \\
	 -\frac{d_2 u_1 \mathcal{Y}_1}{\sqrt{\mathcal{B}\mathcal{C}}} & -\frac{u_2}{\sqrt{\mathcal{C}}} & 
	 \frac{d_3 u_1 \mathcal{Y}_2}{\sqrt{\mathcal{B}\mathcal{C}}}\\
	 \frac{(\mathcal{C} d_3 - d_2 u_2 u_3)\mathcal{Y}_1 \mathcal{Y}_2 }{\sqrt{\mathcal{A}\mathcal{B}\mathcal{C}}} & 
	 \frac{u_1 u_3 \mathcal{Y}_2}{\sqrt{\mathcal{A}\mathcal{C}}} & 
	 \frac{\mathcal{C} d_2 \mathcal{Y}_1^2 + d_3 u_2 u_3 \mathcal{Y}_2^2}{\sqrt{\mathcal{A}\mathcal{B}\mathcal{C}}}
\end{pmatrix} \label{VCKM} \\
\mathcal{A} = \mathcal{C} \mathcal{Y}_1^2 + u_3^2 \mathcal{Y}_2^2 \,, \quad 
\mathcal{B} = d_2^2 \mathcal{Y}_1^2 + d_3^2 \mathcal{Y}_2^2 \,, \quad
\mathcal{C} = u_1^2 + u_2^2 \,, \nonumber \\
		m_\ro{c}^2 = \tfrac{1}{2} \mathcal{Y}_2^2 (u_1^2 + u_2^2 + u_3^2)  \quad 
		m_\ro{t}^2 = \tfrac{1}{2} [\mathcal{Y}_1^2 (u_1^2 + u_2^2) + \mathcal{Y}_2^2 u_3^2 ] \,, \nonumber \\
		m_\ro{s}^2 = \tfrac{1}{6}(d_3 - d_2)^2 \mathcal{Y}_2^2 \,, \quad
		m_\ro{b}^2 = \tfrac{1}{2} (d_2^2\mathcal{Y}_1^2 + d_3^2\mathcal{Y}_2^2) \,, \nonumber
\end{eqnarray}
while the $u$- and $d$-quarks as well as charged leptons and light neutrinos do not acquire 
masses at tree level. In the limit of small $\mathcal{Y}_2/\mathcal{Y}_1 \ll 1$ suggested by 
the second- and third-generation mass hierarchies (\ref{eq:ratio}), an approximate Cabibbo 
mixing arises, with the angle $\theta_{\rm C} \approx \arctan(u_1/u_2)$.
Indeed, in this limit the top-bottom mixing element $V_{\rm tb} \simeq 1 -
(\mathcal{Y}_2/\mathcal{Y}_1)^2 \sim {\cal O}(1)$ is well under control. Moreover, 
the same ratio (\ref{eq:ratio}) provides a strong suppression for $V_\ro{td}$, 
$V_\ro{ts}$, $V_\ro{bu}$ and $V_\ro{bc}$ CKM elements, in agreement with measurements.
In addition, small tree-level contributions to the masses and CKM from a seesaw-type 
mixing with the heavy vector-like quarks are present. This is similar to the model presented in Ref.~\cite{Morais:2020ypd} which we refer to for further details.
Furthermore, loop contributions generate additional (small) terms to the CKM mixing entries and fermion masses. Note, the minimal scenario that is compatible with the CKM quark mixing and mass spectrum corresponds to a Three Higgs Doublet Model with $d_{1,3}=u_3=0$; a suitable benchmark scenario for detailed phenomenological explorations.

Note, as an interesting possibility for future studies, the fields $\bm{\Sigma}, \bm{\Sigma}^\prime$ can break $\E{6}$ not only to the trinification group but also to $\SU{6}{} \times \SU{2}{}$. The two scenarios have an intersection and the presence of adjoint $\bm{{\cal A}}$ which breaks trinification at a lower scale $M_3$ makes a difference in the associated symmetry breaking pattern and in the corresponding low-energy SM-like EFT limit. A further analysis would be necessary to conclude on which of these two $\E{6}$ breaking schemes is favoured by the vacuum structure of the theory, as well as by phenomenology. Such studies should include the full RG flow analysis and the matching to the SM Higgs and flavor sectors in both scenarios.

In summary, the suggested flavored SUSY-GUT framework exhibits two-fold unification in the gauge 
and Yukawa sectors as a consequence of the Higgs-matter and the SM gauge couplings' Grand 
Unification under $\E{6}$. While higher-dimensional $\E{6}$ operators in 
the $\E{6} \times \SU{2}{F} \times \U{F}$ GUT theory generate the necessary splittings in the Yukawa 
and gauge sectors, the gauge couplings' RG flow suggests a strongly-decoupled 
energy scale for the soft-SUSY breaking sector, giving rise to a consistent low-scale SM-like EFT.
The latter exhibits the minimum of three light Higgs doublets for the model to be generically 
compatible with SM quark masses and CKM mixing. The main features of the SM fermion spectra 
such as the observed 
top-charm and bottom-strange mass hierarchies as well as a Cabibbo-type mixing in the quark sector 
are generated already at tree level. Other parameters of the light fermion spectra such as the small 
CKM mixing elements, $u,d$-quark and charged lepton masses, neutrino masses and mixing should be 
established at higher-loop orders via a mixture of different-type seesaw mechanisms which 
is planned for further studies. 
But it is already clear that vast phenomenological prospects 
offered in the proposed framework by a rich scalar, neutrino and heavy vector-like fermion 
sectors as well as by the gauge family interactions can be expected in the reach 
of future collider experiments. One of the remaining key theoretical goals for further studies 
is to explore the potential of the ultimate Left-Right-Color-Family gauge and 
Yukawa couplings' unification through a dynamical origin of $\E{6} \times \SU{2}{F} \times \U{F}$
from yet unknown large gauge group ${\cal G}$ at the $M_{\rm GUT}$ scale.
\vspace*{5mm}

\section*{Acknowledgments}
The authors would like to thank Ivo de Medeiros Varzielas and João Rosa for inspiring discussions at early stages of this work. 
The work of APM has been performed in the framework of COST ActionCA16201 “Unraveling new physics at the LHC through the precision frontier” (PARTICLEFACE). APM is supported by the Center for Research and Development in Mathematics and Applications (CIDMA) through the Portuguese Foundation for Science and Technology (FCT - Fundação para a Ciência e a Tecnologia), references UIDB/04106/2020 and UIDP/04106/2020. APM is also supported by the project PTDC/FIS-PAR/31000/2017, the projects CERN/FIS-PAR/0027/2019, CERN/FISPAR/0002/2017 and by national funds (OE), through FCT, I.P., in the scope of the framework contract foreseen in the numbers 4, 5 and 6 of the article 23, of the Decree-Law 57/2016, of August 29, changed by Law 57/2017, of July 19. APM is also supported by the Enabling Green E-science for the Square Kilometer Array Research Infrastructure (ENGAGESKA), POCI-01-0145-FEDER-022217.~R.P.~is supported in part by the Swedish Research Council grants, contract numbers 621-2013-4287 and 2016-05996, as well as by the European Research Council (ERC) under the European Union's Horizon 2020 research and innovation programme (grant agreement No 668679). W.P.\ has been supported by DFG, project nr.\ PO-1337/7-1.

\bibliographystyle{mdpi}
\bibliography{bib}

\end{document}